# Magnetosheath ion field-aligned anisotropy and implications for ion leakage to the foreshock

Terry Zixu Liu[1], Vassilis Angelopoulos[1], Hui Zhang[2], Andrew Vu[1], and Joachim Raeder[3]

[1]Department of Earth, Planetary, and Space Sciences, University of California, Los Angeles, Los Angeles, CA, USA. [2]Geophysical Institute, University of Alaska Fairbanks, Fairbanks, AK, USA. [3]Institute for the Study of Earth, Oceans, and Space, University of New Hampshire, Durham, USA

**Key Points**

1. Ion field-aligned anisotropy persists across the magnetosheath, caused by a secondary thermal population of magnetosheath ions.

2. Larger IMF strength and solar wind dynamic pressure and/or energy flux favor stronger field-aligned anisotropy.

3. Near the bow shock, the foreshock ion velocity and density are modulated by the magnetosheath ion field-aligned anisotropy.

**Abstract**

   The ion foreshock is highly dynamic, disturbing the bow shock and the magnetosphere-ionosphere system. To forecast foreshock-driven space weather effects, it is necessary to model foreshock ions as a function of upstream shock parameters. Case studies in the accompanying paper show that magnetosheath ions sometimes exhibit strong field-aligned anisotropy towards the upstream direction, which may be responsible for enhancing magnetosheath leakage and therefore foreshock ion density. To understand the conditions leading to such an anisotropy and the potential for enhanced leakage, we perform case studies and a statistical study of magnetosheath and foreshock region data surrounding ~500 THEMIS bow shock crossings. We quantify the anisotropy using the heat flux along the field-aligned direction. We show that the strong field-aligned heat flux persists across the entire magnetosheath from the magnetopause to the bow shock. Ion distribution functions reveal that the strong heat flux is caused by a secondary thermal population. We find that stronger anisotropy events exhibit heat flux preferentially towards the upstream direction near the bow shock and occur under larger IMF strength and larger solar wind dynamic pressure and/or energy flux. Additionally, we show that near the bow shock, magnetosheath leakage is a significant contributor to foreshock ions, and through enhancing the leakage the magnetosheath ion anisotropy can modulate the foreshock ion velocity and density. Our results imply that likely due to field line draping and compression against the magnetopause that leads to a directional mirror force, modeling the foreshock ions necessitates a more global accounting of downstream conditions.

# 1. Introduction

The ion foreshock is filled with backstreaming ions upstream of Earth's bow shock (see review by Eastwood et al., 2005). Due to the interaction between the backstreaming foreshock ions and the solar wind, there are many wave activities and transient structures within the ion foreshock, which can result in disturbances in the magnetosphere and ionosphere (see review by Zhang et al., 2022 and references therein). Some efforts have been recently made to establish predictive models of foreshock transient disturbances with foreshock ion and solar wind parameters as input (Liu et al., 2023a, 2023b, 2023c; Vu et al., 2023). In order to eventually forecast foreshock-driven disturbances, a predictive model of foreshock ions is critically needed.

The origin of foreshock ions can be categorized as solar wind reflection and magnetosheath leakage (e.g., Sonnerup, 1969; Edmiston et al., 1982; Schwartz et al., 1983). Two commonly used reflection models are adiabatic reflection and specular reflection (see review by Burgess et al., 2012 and references therein). The adiabatic reflection occurs when the first adiabatic invariant is conserved, so the solar wind ions reverse their parallel velocity (along the magnetic field) in the de Hoffmann-Teller frame to: $-V_{SWn}/\cos\theta_{Bn}$, where $V_{SWn}$ is the solar wind speed along the bow shock normal. In the specular reflection model, the solar wind ions reverse $V_{SWn}$, so the parallel speed of specularly reflected ions in the shock normal incidence frame is: $-V_{SWn}\cos\theta_{Bn}$. For ions to escape from the bow shock, their parallel velocity projection along the bow-shock-normal component must be larger than the bow-shock-normal component of the $E \times B$ drift velocity. So, the minimum parallel escape speed in the shock normal incidence frame is: $-V_{SWn}\sin^2\theta_{Bn}/\cos\theta_{Bn}$. For specularly reflected ions to stream away from the bow shock, their parallel speed needs to be larger than this minimum escape speed, which requires $\theta_{Bn} < 45°$, a well-known criterion (e.g., Eastwood et al., 2005).

Using observations in comparison with models, Schwartz and Burgess (1984) have shown that both solar wind reflection and magnetosheath leakage take place, but simulations have continued to favor solar wind reflection as the dominant source of the foreshock ions (e.g., Burgess and Luhmann, 1986; Burgess, 1987; Oka et al., 2005). In simulations and models, the magnetosheath ions are commonly considered to be isotropic with respect to the field-aligned direction. Case studies in the accompanying paper (Liu et al., 2023 submitted to JGR) demonstrate that the magnetosheath ions near the bow shock sometimes exhibit strong field-aligned anisotropy towards the upstream direction, which is likely responsible for a significant enhancement of magnetosheath leakage. Thus, the role of magnetosheath leakage in populating the foreshock region may have been previously underestimated.

To further examine the field-aligned anisotropy of magnetosheath ions, we conduct spacecraft observations across the entire magnetosheath from the magnetopause to the bow shock. To determine under what conditions strong field-aligned anisotropy of magnetosheath ions occurs and how the foreshock ion properties could be affected, we perform a statistical study informed by a collection of magnetosheath and foreshock region data surrounding ~500 bow shock crossings. The crossings were selected from four years of observations (2016-2020) by the Time History of Events and Macroscale Interactions during Substorms mission (THEMIS). The dataset and methodology are introduced in Section 2. The case studies are shown in Section 3.1 and the statistical results are shown in Section 3.2. We discuss our results in Section 4 and summarize them in Section 5.

## 2. Data and Methods

We use observations from the Magnetospheric Multiscale (MMS) mission (Burch et al., 2016) and Time History of Events and Macroscale Interactions during Substorms (THEMIS, Angelopoulos, 2008) across the magnetosheath. We analyze plasma data from the MMS fast plasma investigation (FPI) instrument suite (Pollock et al., 2016) and the THEMIS electrostatic analyzer (ESA) (McFadden et al., 2008) and DC magnetic field data from the MMS fluxgate magnetometer (Russell et al., 2016) and the THEMIS fluxgate magnetometer (Auster et al., 2008). We use OMNI data to determine the upstream conditions. With OMNI as input, the Open GGCM model is used to demonstrate global magnetic field geometry. To quantify the directional field-aligned anisotropy of magnetosheath ions, we calculate the magnetosheath ion heat flux projected in the field line direction. The uncertainty of ion heat flux measured by FPI and ESA is $\sim 1 \times 10^9$ eV/cm$^2$/s.

From 2016 to 2020, thanks to their large spacecraft apogee (12-15 $R_E$), the three THEMIS spacecraft frequently crossed the dayside bow shock, with a GSE-Y position mostly within 10 $R_E$. Whenever THEMIS spacecraft is in fast survey mode (with distribution functions at maximum time resolution and appropriate angular resolution), we manually select ~500 bow shock crossings that are associated with foreshock ions (based on ion energy spectra). Our event list can be found in the supporting information (Table S1).

For each crossing, we manually identify downstream and upstream time intervals with relatively stable parameters to avoid the bow shock and any complicated structures around it (discontinuities, foreshock transients, etc.). We then calculate the average plasma and magnetic field parameters within the downstream and upstream time intervals. Because the upstream flow speed is often affected by foreshock ions, we also use the OMNI data to determine the pristine solar wind speed and Mach number.

To obtain the bow shock normal, we use the Merka et al. (2005) model (with upstream parameters as input) and the mixed mode coplanarity method (Schwartz, 1998) (using upstream and downstream parameters). The normal from the coplanarity method often exhibits very tilted direction even when the spacecraft is around the bow shock nose, probably due to measurement uncertainties, shock ripples (e.g., Johlander et al., 2016), reformation (e.g., Burgess, 1989), and other possible deviations from stationary MHD shocks. Because the bow shock is studied in approximately 10 min timescales, rather than using a data-derived localized or transient bow shock normal, we use the model bow shock normal to represent an average normal that is immune to local fluctuations.

## 3. Results

### 3.1. Case Studies

#### 3.1.1. Event 1 – MMS observations

On 2021 January 19, MMS crossed the entire magnetosheath from the magnetopause at [8.4, -2.1, 5.6] $R_E$ in GSE at ~19:43 UT to the bow shock at [10.7, -1.6, 5.9] $R_E$ in GSE at ~21:18 UT (Figure 1). From the magnetopause to the bow shock, the magnetic field strength decreased by ~20 nT, and the $B_x$ component became less negative (Figure 1b). Since the IMF was nearly constant (Figure 1a), the magnetic field variation across the magnetosheath represents spatial variation.

Figure 1f shows that, across the magnetosheath, there was always a parallel heat flux at $\sim 2 \times 10^{10}$ eV/cm²/s and more than $4 \times 10^{10}$ eV/cm²/s near the bow shock (measurement uncertainty is $1 \times 10^9$ eV/cm²/s). In the foreshock, the heat flux was extremely large (out of panel) due to the counter streaming between the foreshock ions and solar wind ions. This suggests that the directional field-aligned anisotropy of magnetosheath ions observed by Liu et al. (2023, submitted to JGR) is not only close to the bow shock; it persists everywhere in the magnetosheath.

To further examine this anisotropy, we show, in Figure 2, ion distribution functions at times corresponding to the vertical dotted lines in Figure 1. In the foreshock (Figure 2a), there were both field-aligned foreshock ions (green and light blue) in the parallel direction and very diffuse, suprathermal foreshock ions (dark blue) in all directions. In the magnetosheath near the bow shock (Figure 2b), diffuse, suprathermal ions (dark blue) transported from the foreshock are seen. The thermal ions (red to green), on the other hand, show very clear asymmetry towards the parallel direction, consistent with observations by Liu et al. (2023, submitted to JGR). Such an anisotropy towards the parallel direction could enhance magnetosheath ion leakage and possibly contribute to the field-aligned foreshock ions observed in Figure 2a.

Deeper in the magnetosheath (Figures 2c – 2e), the thermal ions exhibit enhanced perpendicular anisotropy (typical in the magnetosheath, e.g., see review by Lucek et al., 2005), which weakened the anisotropy in the parallel direction, resulting in smaller parallel heat flux than that near the bow shock (Figure 1f; also see Figure S1 in the supporting information for the perpendicular anisotropy vs. the parallel heat flux). The suprathermal ions show preference towards the anti-parallel direction (Figures 2c and 2d) as they likely originated from upstream diffuse foreshock ions. Near the magnetopause (Figure 2e), the suprathermal ions became isotropic again, with phase space density (PSD) comparable to their counterpart in the magnetosphere (Figure 2f).

To better characterize this directional field-aligned anisotropy, the black line in Figure 2g shows the 1-D cut of ion distribution in Figure 2d at the E×B drift speed along the field-aligned direction (with PSD normalized to the peak value). The distribution can be fitted as the sum of three Maxwellian distributions (red line): (1) a primary thermal population with thermal speed of ~88 km/s; (2) a suprathermal population with peak normalized PSD of ~$1.3 \times 10^{-5}$, thermal speed of ~617 km/s, and bulk speed of ~155 km/s in the anti-parallel direction relative to the primary thermal population (as it originates from upstream diffuse foreshock ions); (3) a secondary thermal population with peak normalized PSD of ~0.01, thermal speed of ~246 km/s, and bulk speed of ~68 km/s in the parallel direction relative to the primary thermal population. The three-Maxwellian distribution fits the data very well, except that it misses some asymmetry of the primary thermal population in the parallel direction (difference between black line and red line at ~100-300 km/s). Both the asymmetry and the secondary thermal population contribute to the observed parallel heat flux.

One possible cause for these parallel populations, in addition to the primary thermal population, could be magnetospheric leakage (e.g., Anagnostopoulos et al., 1986, 2000; Sibeck et al., 1988). To examine this possibility, we compare the normalized PSD in the magnetosphere near the magnetopause (blue line in Figure 2g) with that in the magnetosheath. The PSD of thermal populations in the magnetosphere was several orders of magnitude smaller than that in the magnetosheath, indicating that the magnetospheric leakage should not be the cause. The PSD of suprathermal populations, on the other hand, matches the data well in the anti-parallel direction. If we plot the 1-D PSD corresponding to Figure 2e, the PSD of suprathermal ions matches in both directions. This comparison suggests that there was very likely ion exchange/leakage across the magnetopause for suprathermal populations, but not for thermal populations.

Another possible cause for the additional parallel population could be a mirror force in the magnetosheath (see magnetic field gradients in Figure 1b), which may continuously drive some ions along the field line. To explore the global magnetic field geometry in the magnetosheath, we use the Open GGCM model with OMNI data as input (Figure 3). We present the 3-D simulation results at a planar cross-section that contains the IMF and spacecraft trajectory (simplified as a line in the X direction; Figure 3a) to focus on field lines around MMS on the dawn side. We also present at a planar cross-section that contains the IMF and Sun-Earth line for a more global view (Figure 3b). In these cross-sections, the out-of-plane component of the magnetic field is very small, which allows us to avoid the presentation difficulties caused by projecting 3-D field lines into a 2-D plane.

Along the MMS outbound trajectory, towards the bow shock (red line in Figure 3a), the magnetic field strength decreases (yellow or green to light blue) and the $B_x$ component becomes weaker, consistent with MMS observations (Figure 1b). As already known, such a spatial variation is due to field line draping (e.g., Zwan and Wolf, 1976). Immediately across the bow shock, the magnetosheath field lines follow the Rankine-Hugoniot relations. While approaching the magnetopause, the field lines become more and more parallel to the magnetopause surface. Such a spatial variation from the bow shock to the magnetopause leads to additional compression close to the magnetopause, as commonly observed (e.g., Dimmock et al., 2017)), causing spatially varying magnetic gradient and mirror forces across the entire magnetosheath (see in Figures 3c and 3d the magnetic gradient projection along the magnetic field on the same two planes as Figures 3a and 3b respectively). Such a mirror force could cause some magnetosheath ions to move along the field lines, away from the mirror point, resulting in directional asymmetry/anisotropy.

If this scenario is true, then as the dawn and dusk flanks are on two opposite sides of the mirror point (see red color changing to blue across the magnetopause nose in Figure 3d or also see field lines in Figure 3b), magnetosheath ions on these two sides should experience opposite mirror force directions, leading to opposite field-aligned heat flux. To confirm this scenario, we use conjunction observations by THEMIS and MMS in the magnetosheath on the two sides of the magnetopause nose, as discussed in the next section.

### 3.1.2. Event 2 – MMS and THEMIS conjunction

On 2019 December 17, THEMIS and MMS crossed the magnetopause entering the magnetosheath at GSE-Y = -7.2 $R_E$ and 3-4 $R_E$, respectively (see sketch in Figure 4f; their separation in GSE-Z is only ~2-3 $R_E$). Based on OMNI data (Figure 4a), the IMF was initially in the -X and -Y direction (sketched in Figure 4f) with a weak negative Z component. Later, there was a very thick discontinuity (between two vertical dotted lines), which increased $B_x$ and $B_z$ from negative to positive. We use this $B_z$ sign change as a marker to identify the passage of this thick discontinuity in the magnetosheath by THEMIS (Figure 4b) and MMS (Figure 4d). To avoid complications from the variable magnetic field geometry during the discontinuity, we focus on the time interval before it when both THEMIS and MMS were close to the magnetopause (between vertical dashed line and dotted line).

Before the discontinuity, the approximate field line geometry in the XY plane was qualitatively mirror-symmetric to that in Figure 3. Due to their large separation in GSE-Y direction (>10 $R_E$), the two spacecraft were very likely on the two sides of mirror point (Figure 4f). Figures 4b and 4d show that the two satellites observed opposite signs of the $B_x$ component (see blue arrows in Figure 4f), suggesting convergence and thus compression of the field lines around the magnetopause nose caused by the field line draping. Figures 4c and 4e show that THEMIS observed parallel heat flux, whereas MMS observed anti-parallel heat flux, both directed away from the magnetopause nose, where a compression region, or a mirror point, is therefore expected. Although the time resolution of THEMIS plasma data is low, as the field-aligned heat flux does not show strong sign variation under stable magnetic field geometry as observed by MMS in the two events, THEMIS observations are still suggestive. Overall, we see that the field-aligned heat flux is very sensitive to the magnetic field geometry and its magnetosheath location, in a sense consistent with our suggested scenario. Another THEMIS conjunction observation example can be found in the supporting information (Figure S2)

### 3.2. Statistical Study

Using ~500 THEMIS bow shock crossings, we perform a statistical study of magnetosheath ion field-aligned anisotropy (indicated as field-aligned heat flux) near the bow shock as a function of upstream parameters. We define field-aligned heat flux towards upstream direction to be positive. We also examine the relationship between magnetosheath ion field-aligned heat flux and foreshock ion properties. To calculate the moments of foreshock ions, we remove the solar wind beam from ion distribution functions by setting up a velocity radius around the center of solar wind beam (see details in Liu et al. (2017)).

### 3.2.1. Magnetosheath field-aligned anisotropy

We first determine which upstream parameters affect the magnetosheath field-aligned anisotropy. Figure 5a shows the field-aligned heat flux of magnetosheath ions towards the upstream direction plotted versus $\theta_{Bn}$ (the angle between IMF and bow shock normal). Because we only select bow shock crossings in the presence of foreshock ions, $\theta_{Bn}$ is mostly below 60°. We see that the heat flux is predominantly towards the upstream direction, no matter whether the IMF $B_n$ component is positive (black dots) or negative (blue dots).

Figures 5b, 5c, 5d and 5e show the heat flux plotted versus solar wind speed, density, IMF strength, and Alfvén Mach number, respectively (also see joint probability distributions normalized to the product of marginal probability distributions in Figure S3 in the supporting information). To evaluate their monotonic relationships, we calculate Spearman's rank correlation coefficients and Kendall's rank correlation coefficients (see Table 1). The heat flux shows moderate positive trend with the solar wind speed and the IMF strength, weak positive trend with the solar wind density, and weak negative trend with the Alfvén Mach number. These results are consistent with the trend seen from the median values (red in Figure 5).

However, because the solar wind parameters are not mutually independent, especially fast solar wind has tenuous density, we examine the partial correlation with one parameter fixed at a time (Table 1). We find that when the solar wind density (speed) is fixed, the solar wind speed (density) shows stronger correlation with the heat flux, and both parameters are not affected by the fixed IMF strength. This indicates a possibility that the heat flux could be positively correlated with both the solar wind speed and density, which is weakened by the anti-correlation between the two parameters. It is possible that the heat flux is correlated with $n_{sw} \cdot V_{sw}^a$ where a is an unknown index. We find that when a = 1, the Spearman's coefficient is 0.39. When a = 2 and 3, the coefficient increases to 0.60 and 0.63, respectively. When a > 3, the coefficient decreases (until converging to the correlation with the solar wind speed). Thus, it is very likely that the heat flux is correlated with the solar wind dynamic pressure and/or energy flux (see their scatter plots in Figure S4 in the supporting information). (Their partial correlation coefficient with fixed IMF strength is 0.37 and 0.47, respectively.) This result indicates that the solar wind energy input is very likely the source of the heat flux rather than magnetospheric leakage, consistent with Event 1 in Section 3.1.1.

Because the IMF strength does not have strong correlation with either the solar wind speed or density, its partial correlation does not show large differences from Spearman's correlation. However, the Alfvén Mach number is clearly affected. When the IMF strength is fixed, the negative relationship between the Alfvén Mach number and the heat flux reverses to be positive, indicating that their negative relationship is mostly contributed by the positive correlation with the IMF strength. Although the solar wind speed and density contribute to the positive relationship, the anti-correlation between them weakens it. Also due to this anti-correlation, when either the solar wind speed or density is fixed, the negative relationship becomes weaker. Overall, the negative relationship between the Alfvén Mach number and the heat flux is not likely due to a physical reason but mostly due to the relationship with the IMF strength.

The dependence of heat flux on $\theta_{Bn}$ is not monotonic leading to small correlation coefficients (Figure 5a), but through normalized probability distributions (Figure S3) we can see that the heat flux is more likely larger at $\theta_{Bn} > 30°$ than at $\theta_{Bn} < 30°$. One possible contribution to the heat flux from the shock obliquity, represented by $\theta_{Bn}$, is that foreshock ions are more diffuse at smaller

$\theta_{Bn}$ (see review by Burgess et al., 2012) and thus they can more easily propagate downstream into the magnetosheath (e.g., Karlsson et al., 2021), weakening the anisotropy towards the upstream direction (e.g., Figure 2).

**Table 1.** The Spearman, Kendall, and partial correlation coefficients between the upstream parameters and the field-aligned heat flux.

|  | Spearman | | Kendall | | Partial | | |
| --- | --- | --- | --- | --- | --- | --- | --- |
|  | correlation | significance | correlation | significance | $V_{sw}$ | $n_{sw}$ | $B_t$ |
| $\theta_{Bn}$ | 0.15 | 9e-4 | 0.10 | 9.0e-4 | 0.09 | 0.09 | 0.10 |
| $V_{sw}$ | 0.45 | ~0 | 0.31 | ~0 |  | 0.59 | 0.40 |
| $n_{sw}$ | 0.14 | 0.002 | 0.09 | 0.002 | 0.41 |  | 0.03 |
| $B_t$ | 0.51 | ~0 | 0.35 | ~0 | 0.45 | 0.47 |  |
| $M_A$ | -0.27 | 1e-6 | -0.17 | 1e-6 | -0.06 | -0.13 | 0.15 |

In summary, we find that larger solar wind dynamic pressure and/or energy flux and larger IMF strength favor a larger heat flux, and more oblique bow shock conditions could also contribute but play a minor role. This result is consistent with the proposed scenario is Section 3.1: Large IMF strength and large solar wind dynamic pressure (thus thinner magnetosheath) lead to large magnetic gradient in the magnetosheath favoring ion reflection, and the energy flux of the reflected ions should be proportional to the energy flux of the incoming solar wind ions. Together they contribute to a large heat flux. The role of $\theta_{Bn}$ is likely because the diffuse foreshock ions at small $\theta_{Bn}$ can propagate into the magnetosheath weakening the heat flux in the upstream direction (as shown in Event 1). However, case studies in the accompanying paper (Liu et al., 2023 submitted to JGR) show that under nearly the same values of the aforementioned upstream parameters, the magnetosheath ion field-aligned anisotropy can be quite different. This means that there should be additional parameters controlling the anisotropy, other than the typical upstream parameters considered thus far.

As suggested by case studies in Section 3.1, such parameters could be related to the downstream conditions, such as the magnetic field geometry relative to the magnetopause. It is plausible that the source of anisotropy may be related to the downstream field geometry, specifically how the magnetic field lines are draped in the magnetosheath around the magnetopause and how they map from the near-magnetopause region to the near-bow shock region where the spacecraft observations are made.

To quantify how magnetic field lines are oriented within the magnetosheath and how this orientation results in the above mapping, we calculate the angle ($\zeta$) between the downstream magnetic field and the spacecraft position vector (at the bow shock) in the GSE-YZ plane. When $\zeta$ approaches 90°, the field line is towards the magnetopause flank along the azimuthal direction. When $\zeta$ approaches 0° (or 180°), the field line could map either towards the magnetopause nose along the radial direction or tailward away from the magnetopause nose, largely depending on the sign of $B_x$. We thus include $B_x$ in the definition of $\zeta$ by using its supplement (180° - $\zeta$) when $B_x < 0$. In this definition, when $\zeta \sim 0°$, the field line maps towards the magnetopause nose from the bow shock (e.g., similar to field lines on the dawn side around the bow shock nose in Figures 3a). When $\zeta \sim 180°$, the field line either maps from the bow shock to the tail (e.g., like the field lines around the dusk flank in Figure 3a) or wraps around the magnetopause nose (e.g., like the field line around the dawn flank in Figure 3a). In the accompanying paper with the case studies (Liu et al., 2023 submitted to JGR), the event with (without) strong anisotropy has $\zeta \sim 180°$ (90°).

Figure 6 compares the probability distribution of heat flux normalized to the peak portion at small and large $\zeta$ and the joint probability distribution normalized to the product of marginal probability distributions between ζ and the heat flux. It shows that large heat flux more likely occurs at large ζ. ζ does not show any dependence on $\theta_{Bn}$ or IMF strength (i.e., the local upstream parameters) as it is determined by the spacecraft position and magnetic field geometry relative to the magnetopause (especially the sign of $B_x$). ζ is just a simple example to demonstrate that other than the typical upstream parameters considered earlier, downstream parameters can also affect the magnetosheath ion properties and possibly foreshock ion properties through magnetosheath leakage. It is likely that there are more or better downstream parameters than ζ suggested here. A more global view is needed to model the magnetosheath ion properties and characterize magnetosheath leakage more precisely, as opposed to the present practice of considering the local bow shock as a planar shock.

**3.2.2 Magnetosheath anisotropy effect on foreshock ions**

Next, we statistically examine the origin of foreshock ions and whether the magnetosheath ion anisotropy could play a role in that origin. It is possible that both the solar wind reflection and magnetosheath leakage occur simultaneously, and the bulk velocity of the foreshock ions is determined by the dominant source. Figure 7 compares the foreshock ion parallel speed in the shock normal incidence frame with $-V_{SWn}/\cos\theta_{Bn}$, $-V_{SWn}\cos\theta_{Bn}$, and $-V_{SWn}\sin^2\theta_{Bn}/\cos\theta_{Bn}$, respectively, which (approximately) correspond to adiabatic reflection, specular reflection (only for $\theta_{Bn} < 45°$), and the minimum escape speed from the bow shock. Figures 7a and 5b show that although for some events the parallel speed may follow the two reflection models, the majority of the events have foreshock ion speeds smaller than predicted by those models' speeds. Conversely, Figure 7c shows that for most of the events the foreshock ion parallel speed agrees well with the model escape speed, suggesting that magnetosheath leakage likely dominates the foreshock ion speed. If the foreshock ion speed is dominated by solar wind (adiabatic or specular) reflection, there should be an additional mechanism that decreases that speed to match the observed foreshock ion parallel speed. One possibility for such a decrease is that foreshock ions have become very diffuse (see review by Burgess et al., 2012) and the bulk velocity is filtered by the escape speed. This possibility, however, can only explain a limited number of events, those with small $\theta_{Bn}$, and thus cannot account for the observed inconsistency of the two reflection models with the data. Therefore, magnetosheath leakage is very likely a significant contributor to the foreshock ion population (at least near the bow shock), and the magnetosheath anisotropy towards the upstream direction is expected to enhance such contributions (as already seen from case studies (Liu et al., 2023 submitted to JGR)).

In Figure 7c, there are some scattered deviations between the observed foreshock ion parallel speed and model escape speed ($-V_{SWn} \sin^2 \theta_{Bn} / \cos \theta_{Bn}$). If magnetosheath leakage is the dominant origin of foreshock ions, it is possible that larger magnetosheath ion anisotropy towards the upstream direction could result in larger speed deviations from observations to the model (since more ions are faster than the escape speed). Figure 8a compares the field-aligned heat flux of magnetosheath ions towards the upstream direction with the speed deviations. There is indeed a trend in the data with Spearman's coefficient of 0.30, showing that larger heat flux favors larger speed deviations (although heat flux is not necessarily the best parameter to quantify the leakage, it is still suggestive of it). Because the heat flux is related to the solar wind speed (Figure 5b), we also investigate the dependence between the solar wind speed and speed deviations. Figure S5 (in the supporting information) shows that faster solar wind speed is not correlated with larger speed deviations and thus cannot be the origin of the observed trend in Figure 8a.

Case studies in Liu et al. (2023 submitted in JGR) show that stronger field-aligned anisotropy of magnetosheath ions can result in larger foreshock ion density. Consistent with these case studies,

Figure 6b shows a weak trend that larger heat flux favors larger foreshock ion density normalized to the solar wind density with Spearman's coefficient of 0.24. To examine whether such a trend is caused by the upstream parameters, Figure S6 (in the supporting information) shows that the normalized foreshock ion density does not depend on the solar wind speed and is larger when the IMF strength is smaller and the Alfvén Mach number is larger (e.g., Treumann, 2009). These dependencies on the upstream parameters are mostly opposite to those of the heat flux and thus cannot explain the trend in Figure 8b. In other words, even though the upstream parameters that favor larger heat flux tend to cause smaller normalized foreshock ion density, there is still a positive trend in Figure 8b. We also examine that different $\theta_{Bn}$ ranges do not affect the positive trend in Figure 8b. Therefore, magnetosheath ion heat flux indeed plays a role in also modulating the foreshock ion density.

## 4. Discussion

Why the magnetosheath ions show field-aligned anisotropy is still a puzzle. We propose a scenario that in the magnetosheath near the bow shock there exist both newly heated ions from the solar wind and previously heated ions streaming back towards the upstream region from deeper into the magnetosheath. Thus, this anisotropy is determined by how many previously heated ions stream back up. There could be magnetic mirror forces deeper in the magnetosheath that return some magnetosheath ions back, e.g., because the magnetic field lines are compressed as they drape and piled up at a specific location against the magnetopause (e.g., Dimmock et al., 2020 and Figures 1b and 3). The piling up of such field lines near the magnetopause nose also results in a plasma depletion layer that favors the bulk plasma ions to stream away (e.g., Phan et al., 1994). This scenario could explain the observed dependence of the magnetosheath ion anisotropy on the upstream parameters as discussed in Section 3.2.1. Additionally, due to the curved shape of bow shock and magnetopause, the field line geometry in the magnetosheath can be very complicated (e.g., Karimabadi et al., 2014). The observed field-aligned anisotropy also depends on where the downstream field lines from spacecraft connect to, as suggested in the case studies and Figure 6.

Below we carry out some simple estimates of the proposed magnetosheath ion reflection. Due to the mirror force, magnetosheath ions experience a parallel speed change $m \frac{dV_\parallel}{dt} \sim - \mu \nabla B$, where $\mu$ is first adiabatic invariant. Assuming $\mu$ is conserved, for simplicity, we have $\Delta V_\parallel \sim - \frac{1}{2} V_\perp^2 \left(\frac{\nabla B}{B}\right) \Delta t$. For the reflected ions to be observed, the speed change must be larger than the ion parallel speed, i.e., $V_{\parallel 0} + \Delta V_\parallel < 0$. Thus, $\frac{1}{2} V_\perp^2 \left(\frac{\nabla B}{B}\right) \Delta t > V_{\parallel 0}$. As $\nabla B \sim \Delta B/(V_{\parallel 0} \Delta t)$, we have $\frac{1}{2} V_\perp^2 \left(\frac{\Delta B}{B}\right) > V_{\parallel 0}^2$. Because magnetosheath ions are very hot, $V_{\parallel 0}$ is dominated by the thermal speed $V_{th}$, and thus $V_\perp^2 > \frac{2B}{\Delta B} V_{th}^2$. Based on observations (e.g., Figures 1b), $\frac{B}{\Delta B} > \sim 2$. Therefore, the thermal energy of the reflected ions (or secondary thermal populations) is a few times the thermal energy of the primary thermal populations. (Although $\mu$ is not conserved, the order of magnitude is not likely affected.) This simple estimate is roughly consistent with the fitting results in Figure 2g. Global hybrid simulations and test particle simulations could help confirm and improve this estimate.

Another possible mechanism for the backward moving magnetosheath ions is magnetospheric leakage across the magnetopause due to dayside reconnection, which has been reported at large energies, tends to hundreds of keV (e.g., Anagnostopoulos et al., 1986, 2000; Sibeck et al., 1988). However, the ion distribution comparison (Figure 2g) show that the magnetospheric ions do not contribute to the backward secondary thermal populations, and our results do not show a dependence of magnetosheath ion field-aligned anisotropy on southward IMF $B_z$ (see Figure 1 and

Figure S7 in the supporting information), thus magnetopause reconnection does not appear to play a role here. Future studies including MMS observations of ion compositions could be conducted in the future.

In the statistical study, we examined properties of magnetosheath ions and foreshock ions associated with bow shock crossings, by virtue of our event selection process. As a result, our statistical results are only valid near the bow shock. Deeper in the magnetosheath, perpendicular anisotropy dominates and weakens the field-aligned anisotropy (e.g., Figure S1 in the supporting information). Farther upstream from the bow shock, the bulk velocity of foreshock ions may be more dominated by the solar wind reflection (especially adiabatic reflection) partly due to time-of-flight effect. Multi-point observations should be conducted to further investigate them.

Shock theories and models usually consider planar shocks with uniform downstream conditions. The bow shock, however, is more complicated, due to its curved shape following the curved magnetopause. Although the bow shock itself can be locally treated as a planar shock, the magnetosheath conditions along the shock normal are not uniform. For example, the magnetic field lines deeper in the magnetosheath are connected to a more distant part of the bow shock and are more bent by the magnetosheath flow along the magnetopause surface. The subsolar magnetosheath flow speed decreases linearly to zero from the bow shock to the magnetopause (e.g., Sibeck et al., 2022), unlike what is expected at ideal planar shocks. These complicated downstream conditions can, in turn, affect the bow shock and foreshock properties. For example, electrons can bounce back-and-forth between two shock-related mirror points within the magnetosheath as magnetic field lines could be connected to the bow shock at two dayside locations, thus causing field-aligned anisotropy (e.g., Mitchell et al., 2012; Mitchell and Schwartz, 2014). A flux transfer event at the magnetopause could create a bow wave that can modify the local bow shock shape resulting in a local foreshock (Pfau-Kempf et al., 2016). Given these examples, ideal shock theories are insufficient to model the bow shock and foreshock. Consideration of the global geometry of the bow shock and foreshock is needed to properly address these questions.

## 5. Summary and Conclusions

In summary, we find that under stable IMF, magnetosheath ion directional field-aligned anisotropy (well-described as the field-aligned heat flux) persists across the entire magnetosheath. By fitting the observed distributions using three Maxwellian distributions, we find that this anisotropy was due to a secondary thermal population moving in the field-aligned direction. Using conjunction observations, we show that this anisotropy heavily depends on the magnetic field geometry and magnetosheath location. Based on these results, we suggest that as magnetic field lines drape around the magnetopause, magnetic field gradients build up throughout the magnetosheath. These gradients, in turn, should reflect some magnetosheath ions away from the maximum compression region (e.g., around the magnetopause nose), leading to a directional asymmetry, or anisotropy. In the future, global hybrid simulations can further examine the proposed scenario under a variety of upstream conditions.

. Our statistical results show that near the bow shock, the magnetosheath ions exhibit field-aligned heat flux nearly always towards the upstream direction. Larger solar wind energy flux and/or dynamic pressure and larger IMF strength favor larger heat flux, and large $\theta_{Bn}$ play a minor role. Other than upstream parameters, downstream parameters such as the magnetic field geometry relative to the spacecraft and magnetopause also affect the anisotropy. We also show that the foreshock ion velocity near the bow shock is well correlated with the model escape speed, suggesting magnetosheath leakage as a significant contributor. Larger magnetosheath ion heat flux towards the upstream direction favors larger deviation of observed foreshock ion parallel speed from the model escape speed and favors larger foreshock ion density normalized to the solar wind density. Our results suggest that other than the typical upstream parameters, magnetosheath ion properties that arise from non-planar shock conditions can modulate the foreshock ion properties. Thus, to model the foreshock ions, magnetosheath ion properties such as field-aligned anisotropy must be considered.

**Acknowledgement**

T. Z. L. is partially supported by NSF award AGS-1941012/2210319, NSF award AGS-2247760, NASA grant 80NSSC21K1437/80NSSC22K0791 and NASA grant 80NSSC23K0086. H. Z. is partially supported by NSF AGS-1352669. We acknowledge support by the NASA THEMIS contract NAS5-02099. We thank K. H. Glassmeier, U. Auster and W. Baumjohann for the use of the THEMIS/FGM data provided under the lead of the Technical University of Braunschweig and with financial support through the German Ministry for Economy and Technology and the German Center for Aviation and Space (DLR) under contract 50 OC 0302. We also thank the late C. W. Carlson and J. P. McFadden for use of THEMIS/ESA data. We also thank the SPEDAS team and the NASA Coordinated Data Analysis Web. Simulation results have been provided by the Community Coordinated Modeling Center (CCMC) at Goddard Space Flight Center through their publicly available simulation services (https://ccmc.gsfc.nasa.gov). The Open GGCM Model was developed by the Joachim Raeder at University of New Hampshire.

**Data availability statement**

THEMIS dataset and OMNI dataset are available at NASA's Coordinated Data Analysis Web (CDAWeb, http://cdaweb.gsfc.nasa.gov/). MMS dataset is available at

https://lasp.colorado.edu/mms/sdc/public/. The SPEDAS software (see Angelopoulos et al. (2019)) is available at http://themis.ssl.berkeley.edu. The event list can be found in Table S1 in the supporting information.

**Figures**

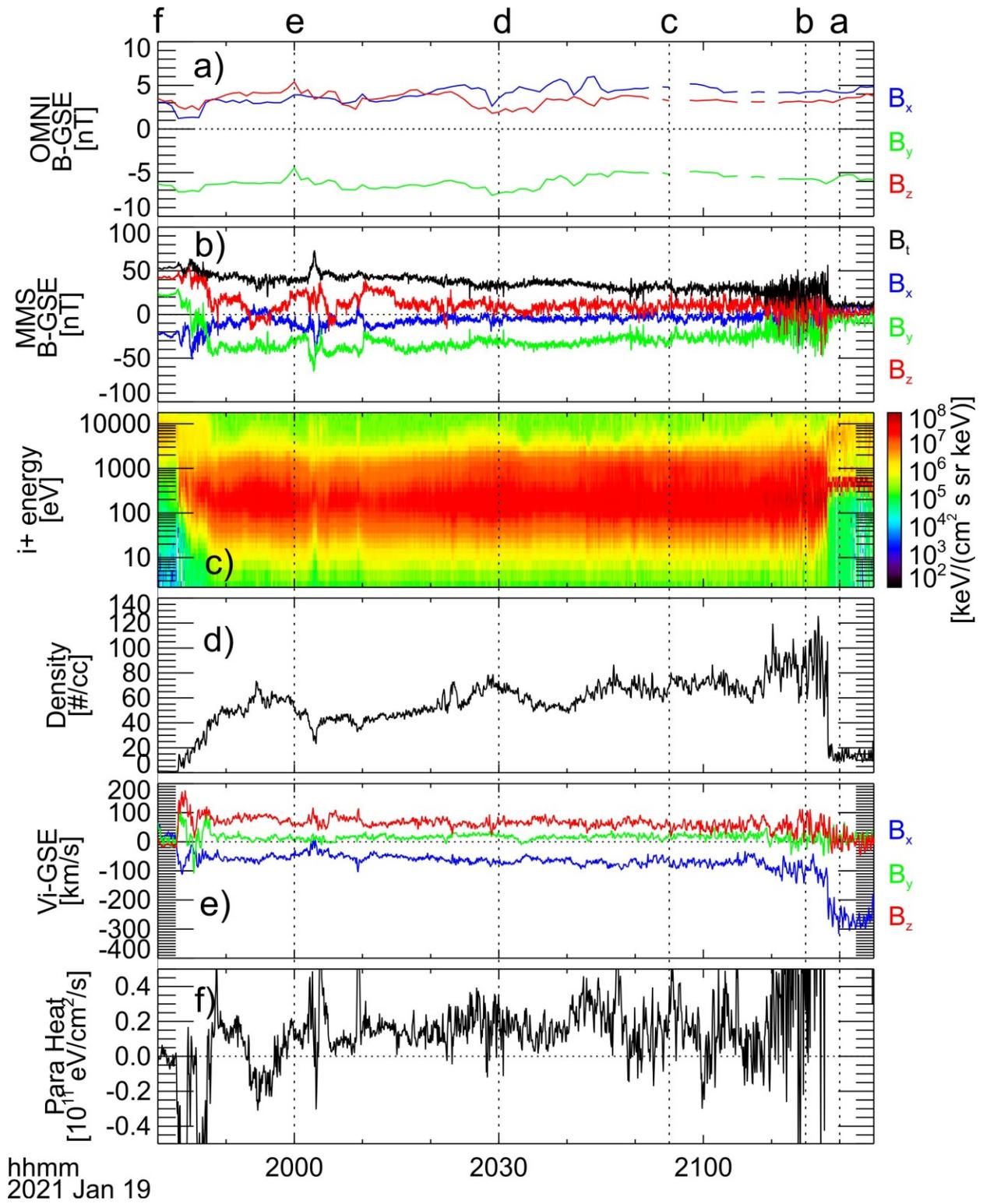

**Figure 1.** MMS1 observations of magnetosheath crossing from the magnetopause to the bow shock. From top to bottom: (a) OMNI magnetic field in GSE, (b) MMS observations of magnetic field in GSE, (c) ion energy spectrum, (d) ion density, (e) ion bulk velocity in GSE, and (f) field-aligned heat flux. Vertical dotted lines and characters above correspond to ion distributions in Figure 2 (vertical Line f is at the left edge of the plot).

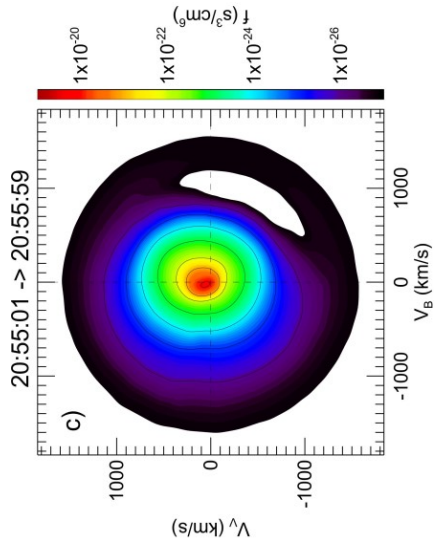
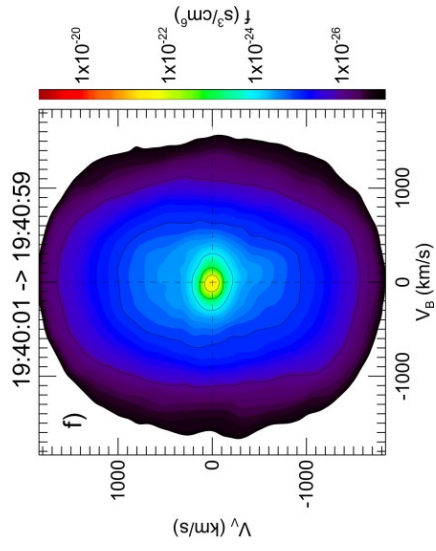
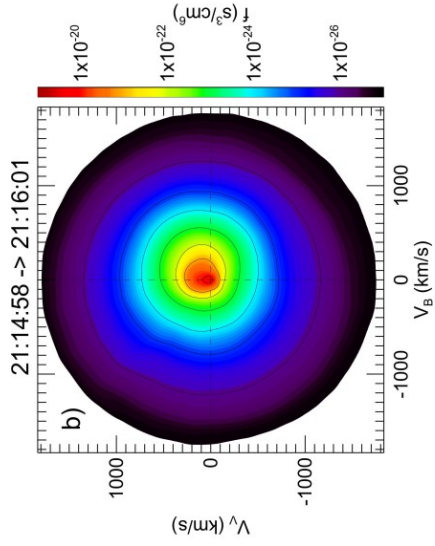
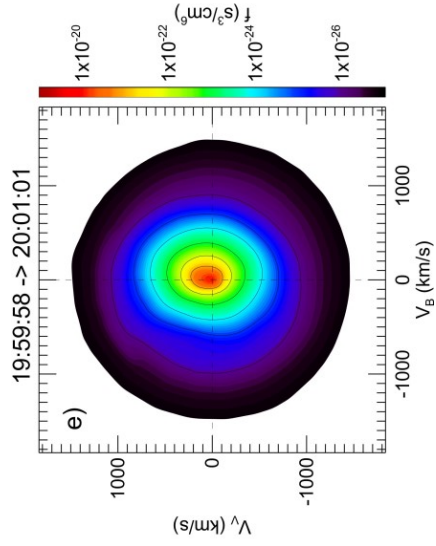
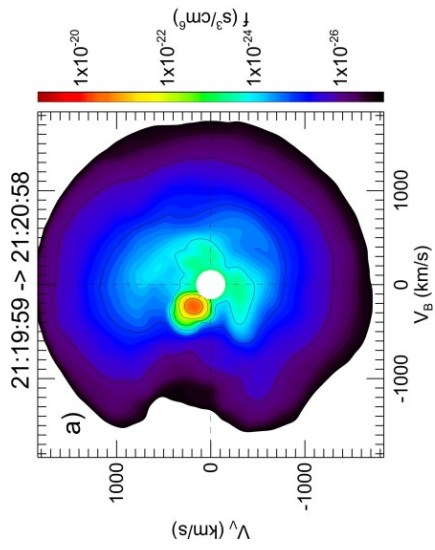
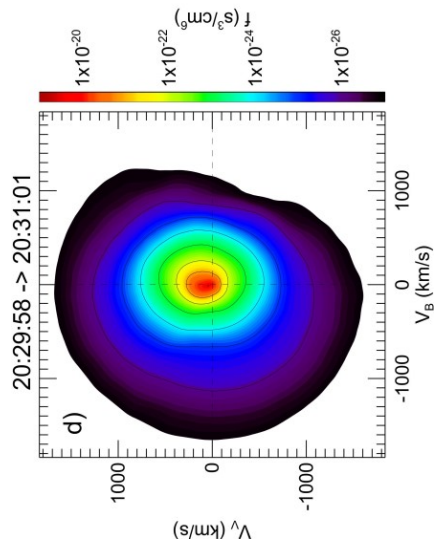

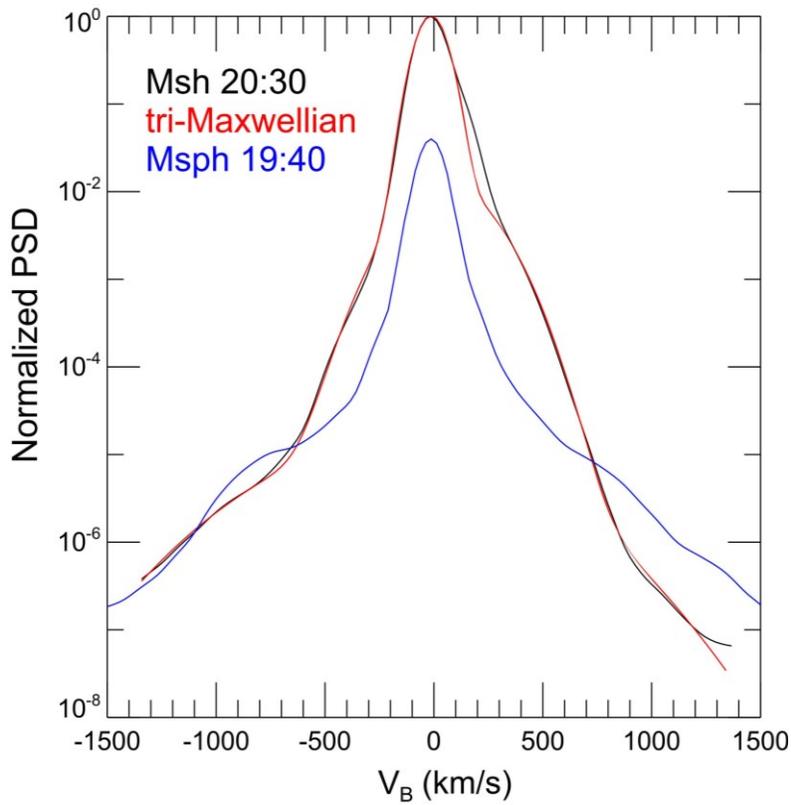

**Figure 2.** Ion distribution functions (averaged over 1 min) corresponding to vertical dotted lines in Figure 1. Their horizontal axis is along the magnetic field and the plane contains the bulk velocity. Panel (g) shows the 1-D distributions from Panels (d) and (f) cut at the E×B drift speed along the magnetic field direction (black and blue, respectively). The PSD is normalized to the peak value. The red line is the fitted three-Maxwellian distribution.

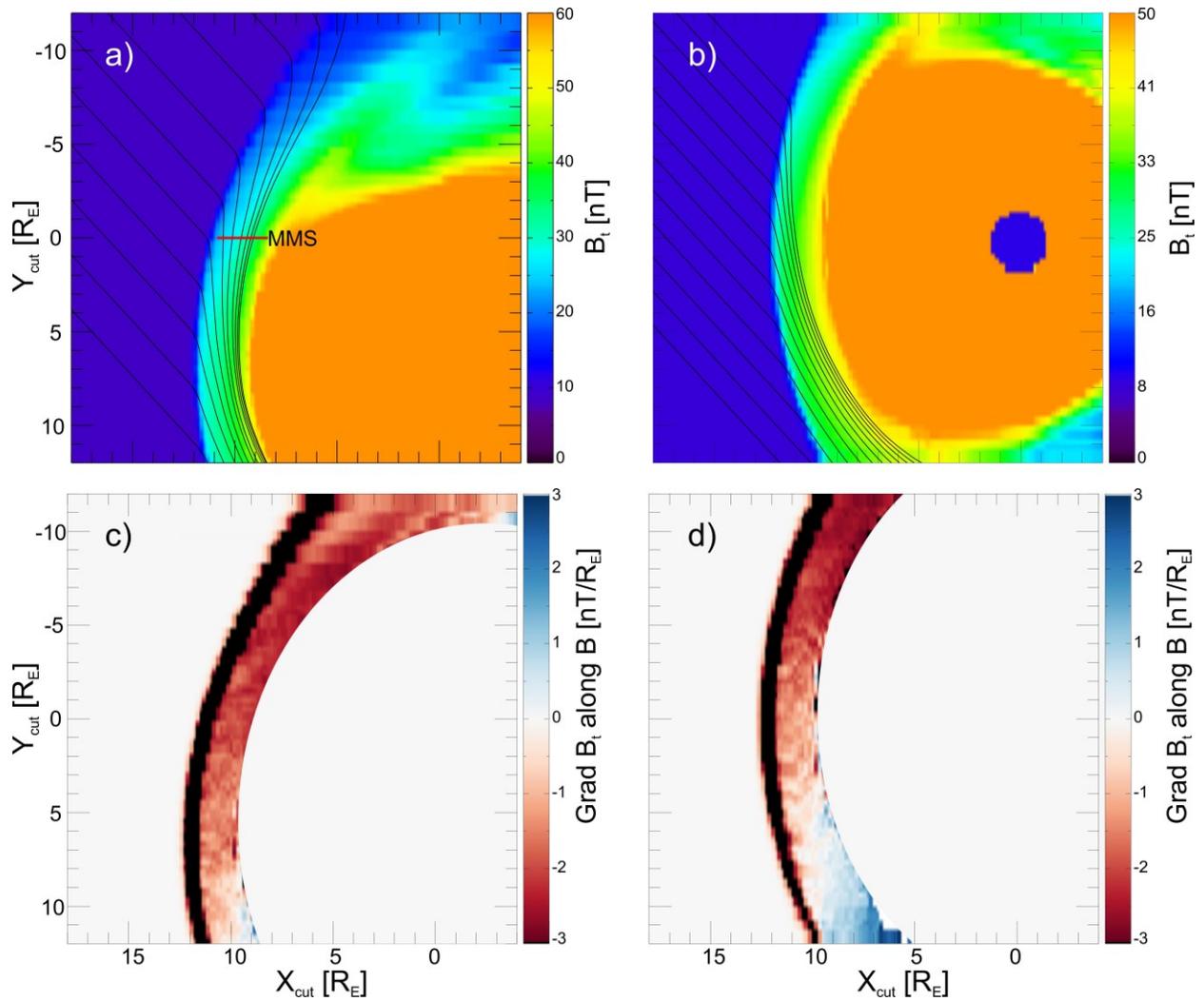

**Figure 3.** The magnetic field geometry from the Open GGCM model. Panels (a) and (b) show the field strength and magnetic field lines in the planes containing the IMF and approximate MMS trajectory (horizontal red line), and containing the IMF and Sun-Earth line, respectively. The origin of vertical axis is defined by the MMS position and Sun-Earth line, respectively. Panels (c) and (d) show the magnetic field gradients projected along the magnetic field in the same two planes as in Panels (a) and (b). The magnetosphere in Panels (c) and (d) is covered to mask the complicated but irrelevant gradients there.

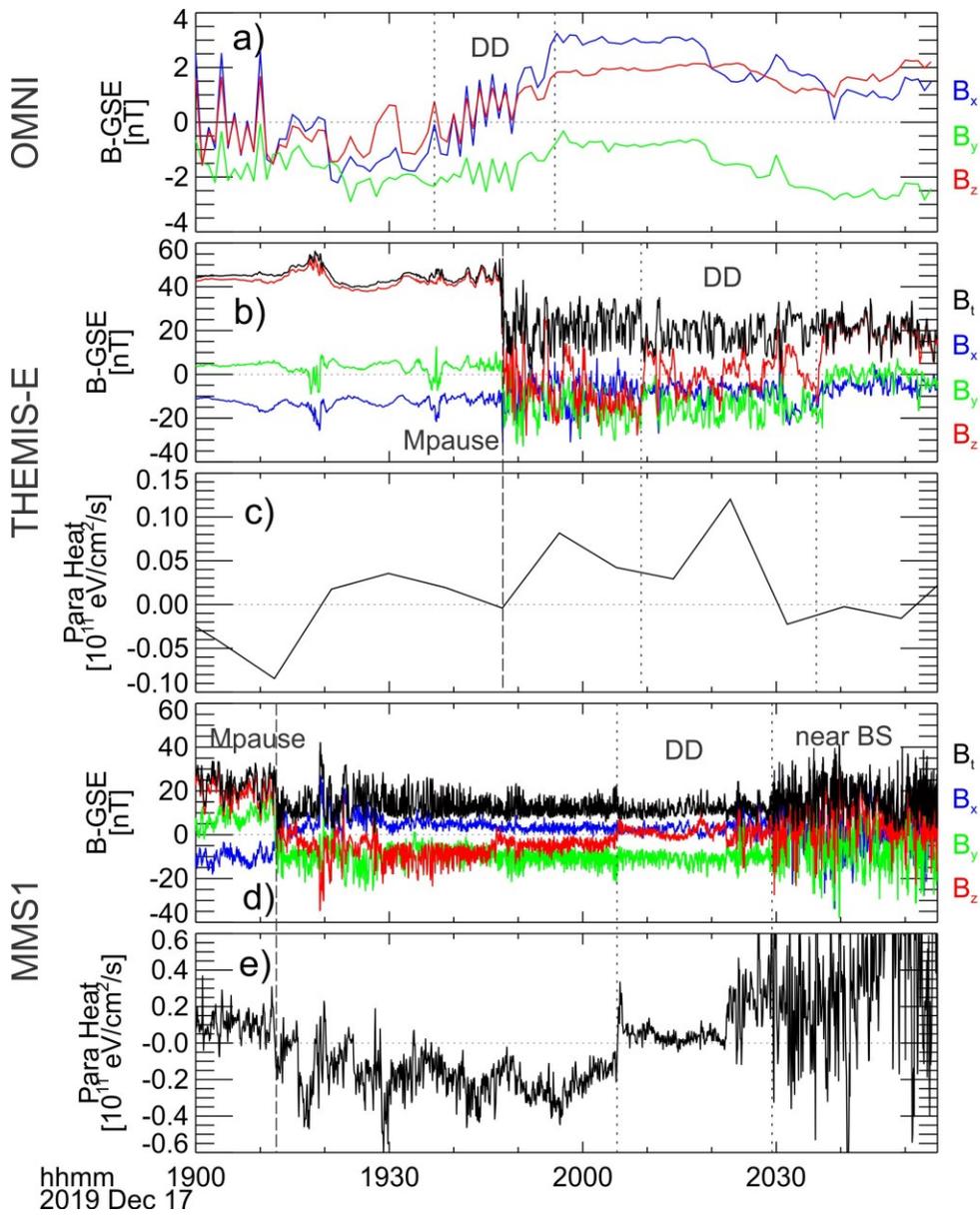

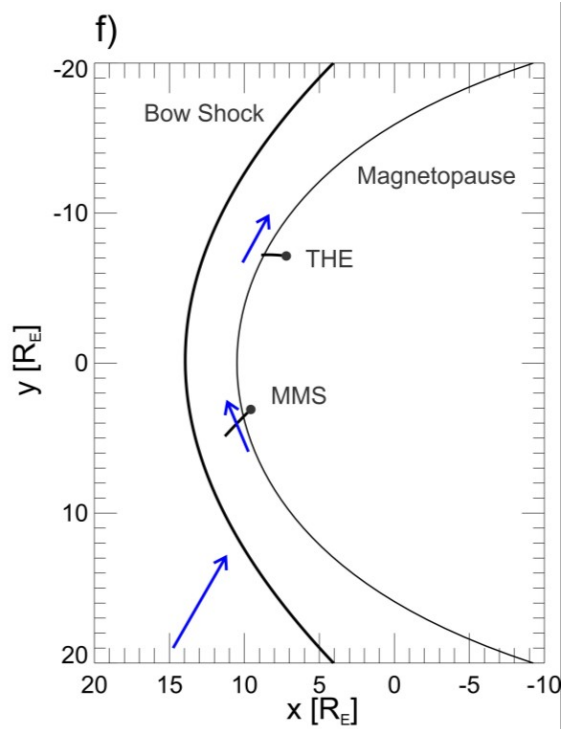

**Figure 4.** MMS-THEMIS conjunction observations (in GSE). (a)-(e) are: OMNI observations of magnetic field, TH-E observations of magnetic field and field-aligned heat flux, and MMS observations of magnetic field and field-aligned heat flux. Vertical dotted lines indicate a thick directional discontinuity (DD), and vertical dashed lines indicate the magnetopause. (f) The geometry of two spacecraft trajectories (black) and the observed magnetic field directions (blue arrows) before the DD encounter are projected in the GSE-XY plane. Black dots indicate the initial spacecraft positions. The Merka et al. (2005) bow shock model and Shue et al. (1988) magnetopause model are used here.

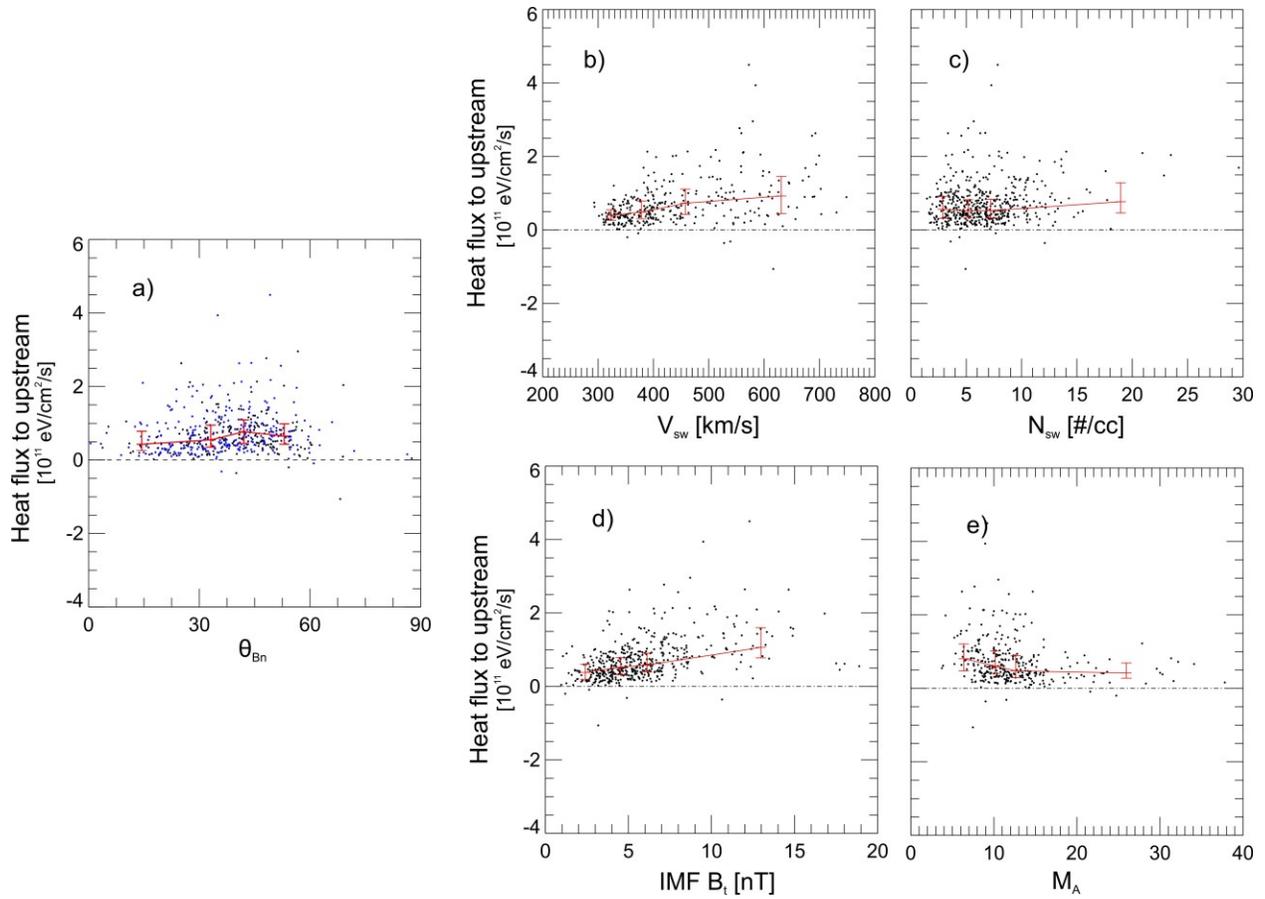

**Figure 5.** The magnetosheath ion field-aligned heat flux towards upstream direction vs. $\theta_{Bn}$, solar wind speed, density, IMF strength, and Alfvén Mach number, respectively. The red bars indicate the median value, lower quartile, and upper quartile. Blue dots and black dots in panel (a) indicate events with negative and positive IMF along the bow shock normal, respectively.

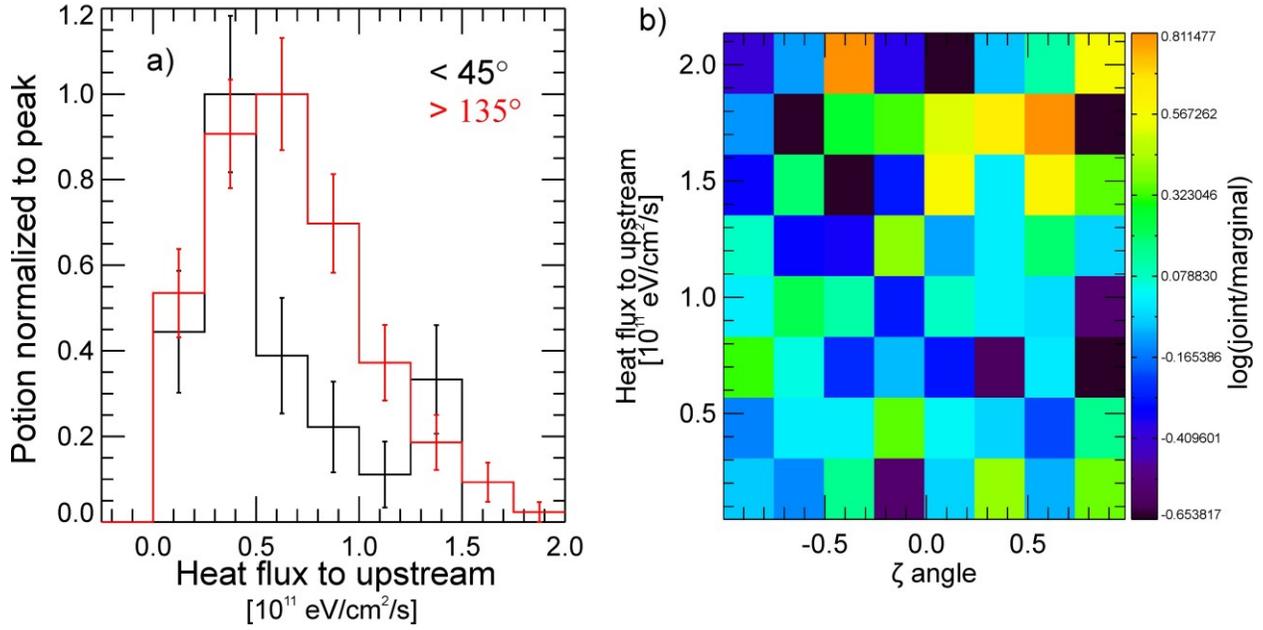

**Figure 6.** (a) The probability distributions of heat flux normalized to the peak portion, at ζ smaller than 45° (black) and larger than 135° (red). The error bars are calculated through binomial distribution. (b) The joint probability distribution normalized to the product of marginal probability distributions between the heat flux and ζ angle.

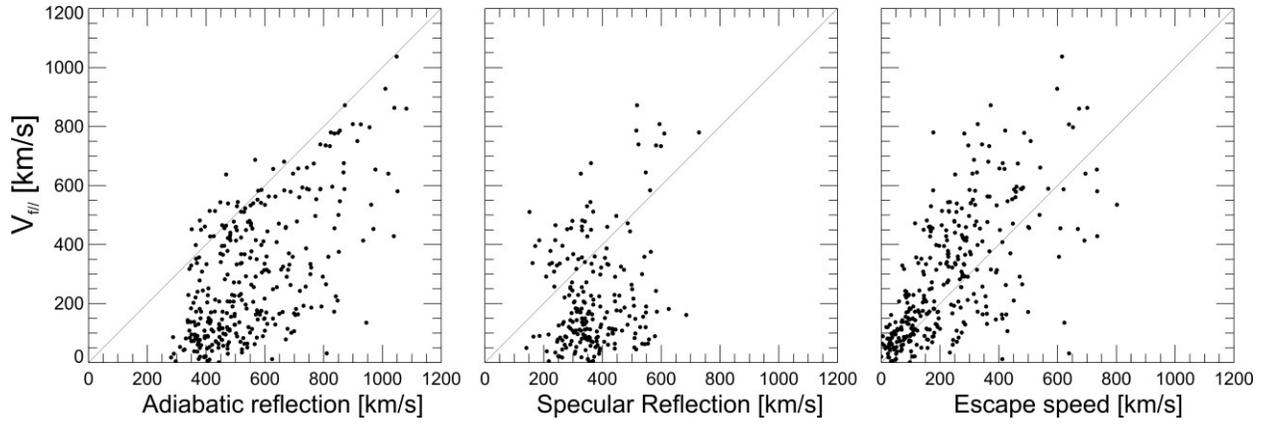

**Figure 7.** The foreshock ion parallel speed in the shock normal incidence frame vs. $-V_{SWn}/\cos\theta_{Bn}$, $-V_{SWn}\cos\theta_{Bn}$, and $-V_{SWn}\sin^2\theta_{Bn}/\cos\theta_{Bn}$, respectively, which approximately represent adiabatic reflection speed, specular reflection speed (only for $\theta_{Bn} < 45°$), and escape speed from the bow shock.

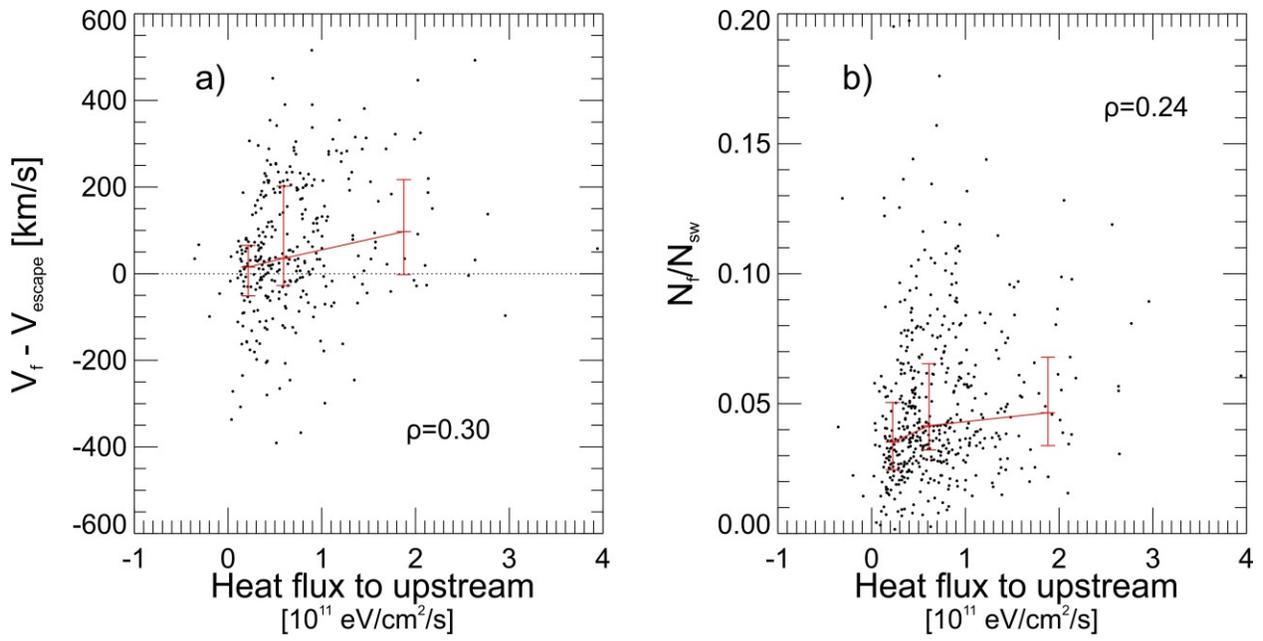

**Figure 8.** (a) The heat flux towards upstream direction of magnetosheath ions vs. the differences between foreshock ion parallel speed and model escape speed and (b) vs. the foreshock ion density normalized to the solar wind ion density. The Spearman's coefficient ρ is shown in each panel.